\documentclass[fleqn,10pt]{SelfArx} 

\definecolor{color1}{RGB}{0,0,90} 
\definecolor{color2}{RGB}{0,20,20} 

\usepackage{hyperref} 
\hypersetup{hidelinks,colorlinks,breaklinks=true,urlcolor=color2,citecolor=color1,linkcolor=color1,bookmarksopen=false,pdftitle={Title},pdfauthor={Author},urlcolor=blue}

\usepackage{graphicx}
\usepackage{natbib}
\usepackage{amsmath}
\usepackage{multirow}
\usepackage{subfigure}

\JournalInfo{Published in Journal of the European Ceramic Society, Vol. 34 (12), 3019–3025, 2014} 
\Archive{\href{http://dx.doi.org/10.1016/j.jeurceramsoc.2014.04.029}{DOI: 10.1016/j.jeurceramsoc.2014.04.029}} 

\PaperTitle{In situ characterization of delamination and crack growth of a CGO-LSM multi-layer ceramic sample investigated by X-ray tomographic microscopy} 

\Authors{R. Bj\o{}rk$^1$, V. Esposito$^1$, E. M. Lauridsen$^1$, P. S. J\o{}rgensen$^1$, J. L. Fife$^2$, K. B. Andersen$^1$, S. P. V. Foghmoes$^1$ and N. Pryds$^1$} 
\affiliation{\textsuperscript{1}\textit{Department of Energy Conversion and Storage, Technical University of Denmark - DTU, Frederiksborgvej 399, DK-4000 Roskilde, Denmark}} 
\affiliation{\textsuperscript{2}\textit{Swiss Light Source, Paul Scherrer Institut (PSI), 5232 Villigen PSI, Switzerland}} 
\affiliation{*\textbf{Corresponding author}: rabj@dtu.dk} 

\newcommand{\keywordname}{Keywords} 

\Abstract{The densification, delamination and crack growth behavior in a Ce$_{0.9}$Gd$_{0.1}$O$_{1.95}$ (CGO) and (La$_{0.85}$Sr$_{0.15})_{0.9}$MnO$_{3}$ (LSM) multi-layer ceramic sample was studied using in situ X-ray tomographic microscopy (microtomography), to investigate the critical dynamics of crack propagation and delamination in a multilayered sample. Naturally occurring defects, caused by the sample preparation process, are shown not to be critical in sample degradation. Instead defects are nucleated during the debinding step. Crack growth is significantly faster along the material layers than perpendicular to them, and crack growth and delamination only accelerates when sintering occurs.}

\begin{document}

\flushbottom 

\maketitle 


\thispagestyle{empty} 

\section{Introduction}
The sintering process is important in a number of technological applications, which utilizes the changes that a powder compact undergoes during sintering. These changes can result in mechanical strengthening, impervious gastight microstructures and improved conductivity of the sintered powder compact. In order to make devices with an enhanced functionality, several different materials must be co-fired, i.e. sintered together. This process can cause internal stresses to appear inside the sample, leading to constrained sintering of the different layers. Although constrained sintering has been investigated in some detail (for a review see e.g. Green et al.\cite{Green_2008}), this is usually done using a continuum theory approach where the samples are characterized on a macroscopic level by measuring the dilatometry of a given sample during sintering. Microscopic properties such as grain and pore size are determined through averaging two dimensional microscopy images of a bisected sample. While these approaches are valid in the case of a homogeneous and isotropic sample, they are not ideal to characterize a sample which contains initial inhomogeneities or defects, as these can develop during sintering. The presence of defects in a sample is critical and fragile fracture is facilitated by the presence of flaws since they induce a concentration of the stress to critical levels.

The fabrication of multilayer components usually is critical because of the constrained sintering conditions, presence of porosity, and defects at the interfaces due to the colamination process of the multilayers. The latter is particularly critical because it is known to lead to delamination. The delamination and crack propagation and how this is correlated to sintering and densification is the purpose of the investigation conducted here. A sample consisting of multiple materials will sinter differently compared to single materials.

Here we examine the influence and evolution of defects on the sintering of a multilayered ceramic sample, with technological relevant properties. In such multilayered ceramic components delamination is the most common mode of failure and it is often associated with the presence of defects at the layer interfaces. In order to follow the evolution of a sample in situ in a non-destructive way, and to obtain information about the internal structure of the sample, X-ray tomographic microscopy (sometime referred to as microtomography) was used. This technique detects the residual energy of a beam of X-rays that passes through the sample. The sample is rotated while a large number of projections is captured and the spatial distribution of the linear attenuation coefficient within the specimen is captured and can then be reconstructed. Because the investigation of the sample is non-destructive, the evolution of a single sample can be followed throughout time. Moreover, the evolution of several and diverse flaws can be followed investigating their reciprocal influence. X-ray tomographic microscopy can be utilized to characterize in situ evolutions of defects during thermal treatments of high temperatures processes, especially when other conventional characterization techniques cannot be used. Particularly, growth of flaws in samples heated during sintering can be monitored in situ while the sintering evolution occurs.

Using X-ray tomographic microscopy to study sintering is widely reported in literature. The method was first performed by Bernard et al.\cite{Bernard_2005}, who studied the sintering of glass powder and lithium borate powder. Almost simultaneously the in situ microstructural evolution of metal powders during sintering using tomography was observed and subsequently modelled for loosely packed copper powder and compacted steel powder \citep{Lame_2004,Tikare_2010}. Further sintering studies using tomography were done by Lautensack et al.\cite{Lautensack_2006}, Schoenberg et al.\cite{Schoenberg_2006}, N\"othe et al.\cite{Nothe_2007}, Vagnon et al.\cite{Vagnon_2008} and Olmos et al.\cite{Olmos_2009}. The rotation of copper particles during free sintering have also been studied \citep{Kieback_2010}. The sintering of glass on a rigid substrate has also been investigated using tomography \citep{Bernard_2011}. Finally, synchrotron X-ray nanotomography has been used to study microstructural changes during sintering of a multi-layer ceramic capacitor, where constrained sintering was shown to results in discontinuous interfaces between the layers \citep{Yan_2012,Yan_2013}.

For almost all the systems studied so far the particle size has been large and only rarely have complete samples been studied. In Vagnon et al.\cite{Vagnon_2008} the particles had diameters between 40 and 60 $\mu$m, and a few hundred particles were followed. In Lame et al.\cite{Lame_2004} the copper particles had the same diameters and the investigated iron powder had particles with sizes between 45 and 150 $\mu$m. In Lautensack et al.\cite{Lautensack_2006} the particle were slightly larger, with diameters equal to 100-120 $\mu$m, but here 10000 particles were followed. In Olmos et al.\cite{Olmos_2009} the largest particles were 63 $\mu$m. In both Bernard et al.\cite{Bernard_2005} and Vagnon et al.\cite{Vagnon_2008} the particles were only tracked in a volume considerably smaller than the complete sample volume. Only in the case of Yan et al.\cite{Yan_2012,Yan_2013} have nano-sized powders been used.

For the systems studied so far, the particles have mostly been copper, which does have some practical applications, but has been used in experiments because the particle size and sintering temperature are ideal for many tomography techniques. Moreover the grain size is not realistic in ceramic technologies where fine powders are preferred as starting materials, as these are known to have superior technological properties \citep{Rahaman_2008}. Also, the sintering behavior of constrained systems on a sample scale has not been studied using tomography.

In this work, we report the sintering behavior of a system with the multiple technological relevant materials Ce$_{0.9}$Gd$_{0.1}$O$_{1.95}$ (CGO) and (La$_{0.85}$Sr$_{0.15}$)$_{0.9}$MnO$_{3 }$ (LSM) \citep{Skovgaard_2012, Rembelski_2012}. The sample have a complex architecture of several alternating layers and is a study-case for gas purification technology \citep{Andersen_2013}. The individual grains are too small to resolve using tomography, however, the multilayer architecture has geometrical features which can be easily investigated using this technique. The sample geometry was slightly modified from the flue gas purification design to have a suitable geometry for the experiment. The alternating layers consisted of thick porous LSM layers (electrodes) and thin dense CGO layers (electrolyte). Although the materials present different comparable themo-mechanical properties (thermal expansion coefficient (TEC)) the thin and dense/thick and porous architecture fabricated by tape-casting is critical. This combination of layers was chosen to characterize critical stress factors which lead to observation of typical crack formation and delamination during firing.

\section{Experimental setup}
The sample examined consisted of alternating layers of two ceramics, Ce$_{0.9}$Gd$_{0.1}$O$_{1.95}$ (CGO, Rhodia, France) and \\(La$_{0.85}$Sr$_{0.15}$)$_{0.9}$MnO$_{3 }$ (LSM,  Haldor Tops\o{}e A/S, Denmark). The LSM powder was calcined at 800$^\circ$C and 1200$^\circ$C and mixed in the ratio 1:1. The LSM tape was added CGO at the ratio 65:35 w/w\%. The CGO tape consists only of CGO powder. As pore former two different types of graphite (D50 of 2.6 $\mu$m and D50 of 10 $\mu$m (both flakes), both Alfa Aesar) were used. The particle size distribution (PSD) for the slurries was measured using a laser diffraction particle size analyser (LS13320, Beckman Coulter, USA). The D50 for the LSM slurry was 1.4 $\mu$m while it was 0.2 $\mu$m for the CGO. Using a so-called doctor blade to control the thickness, the slurries are applied from a vessel onto a moving substrate. The LSM tape was casted at a 1 mm gab and the CGO was casted at a 0.2 mm gab, producing the thick and thin layers. After drying, the tapes were laminated together using the method described by Larsen and Brodersen\cite{Larsen_2008}. The CGO layers have a different absorption of X-rays than the LSM layers, allowing the materials to be distinguished using tomography. The sample investigated using tomography was extracted from the tapecasted multilayer using a 1.6 mm syringe stamped through the sample, after which the cylinder-shaped sample was removed by compress air.

\begin{figure}[t]
  \centering
  \includegraphics[width=1\columnwidth]{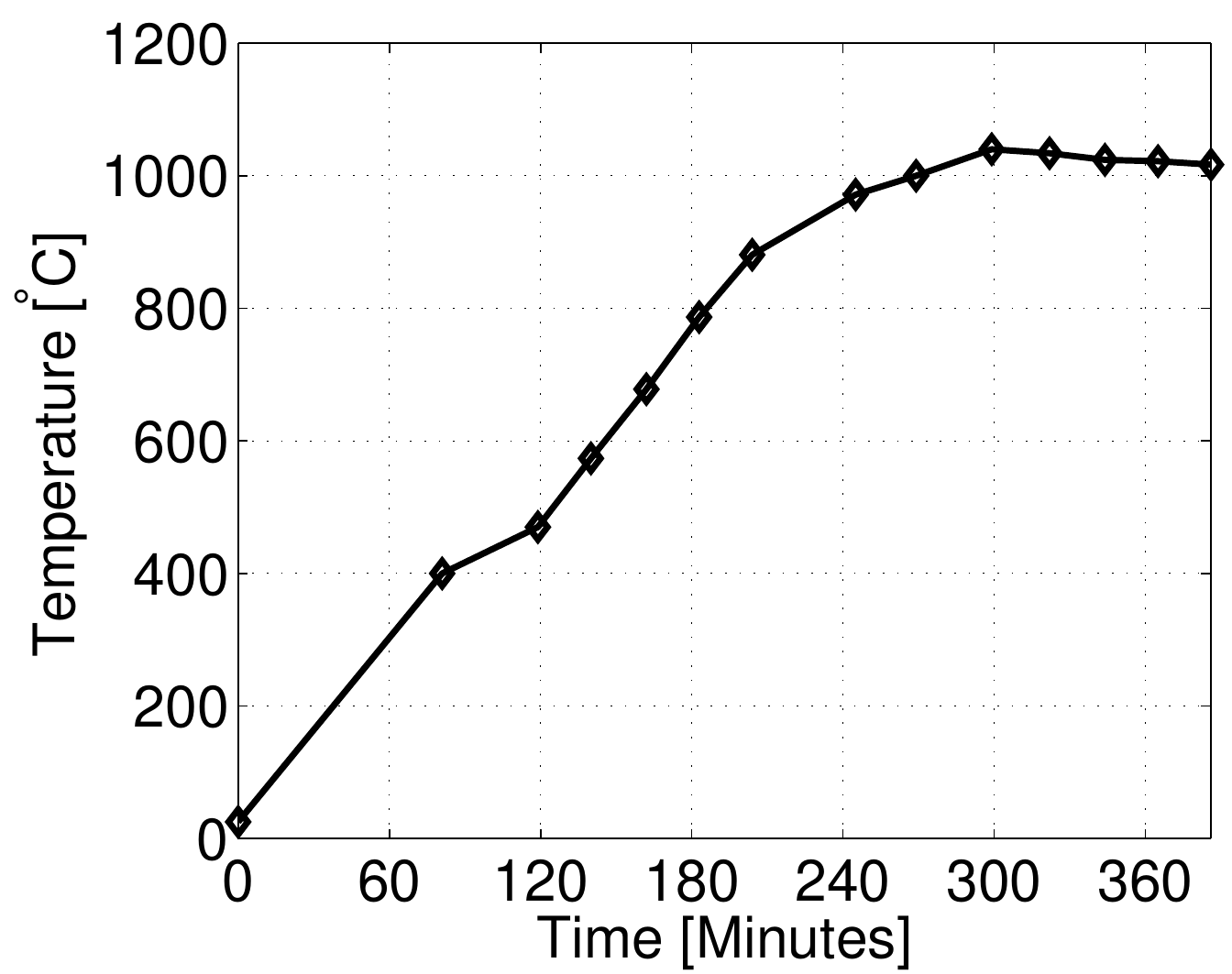}
  \caption{The temperature profile used in the sintering experiment.}
  \label{Fig_Temperature_profile}
\end{figure}

A custom built furnace was used to heat the sample in situ during the tomography experiment. The cylinder-shaped furnace has two holes on opposite sides to allow the X-ray beam to pass through the furnace. A bottom hole allows for sample introduction. The sample was sintered in air and at a temperature of 1000-1050$^\circ{}$C, which is the maximum operating temperature of the furnace. The complete temperature profile recorded during the experiment is shown in Fig. \ref{Fig_Temperature_profile}. Each point on the curve indicates a tomography scan, of which there were 14. The binder material was first burned off at 400$^\circ$C and afterwards the temperature was increased with a ramp rate of 5$^\circ{}$C/min until 900$^\circ{}$C at which the ramp rate was lowered because of furnace limitations. Finally, the temperature was kept isothermal at 1000-1050$^\circ{}$C for about two hours.

The sample was investigated at the TOmographic Microscopy and Coherent rAdiology experimenTs (TOMCAT) beamline at the Swiss Light Source, Paul Scherrer Institut (Villigen, Switzerland) \citep{Stampanoni_2006}. The X-ray photon energy used was 40 keV, with an exposure time of 1100 ms, capturing 1001 projections during 180$^\circ$ rotation of the sample. The detector utilized a Ce-doped YAG scintillator with a thickness of 20 $\mu$m, and a $2048\times{}2048$ pixel CCD camera with a 280c digital/analog converter and a 10 MHz read-out speed. The voxel size was 1.48$\times$1.48$\times$1.48 $\mu$m$^3$.

The reconstructed three-dimensional data from the 14 tomographic microscopy scans are not oriented similarly in space, as both the sample and the sample holder rod moved during the experiment due to thermal expansion. The datasets have therefore been aligned using the Avizo Fire 7.00 software. The data are rotated and translated to fit a single dataset which was chosen to be the reference. This dataset was first aligned such that the layers of CGO were horizontal. This was done using an artificial dataset that contained completely horizontal layers. After aligning all the datasets to the reference dataset, the bottom CGO layer was extracted from all aligned datasets and afterwards aligned again to the bottom CGO layer of the reference dataset. In this way the orientation of the sample in each dataset was first aligned using all layers and afterwards fine-tuned using the bottom CGO layer of the sample. After this procedure all datasets are oriented vertically, such that the CGO layers are horizontal.

In our case, the tomography technique combined with the substantial heat treatment of the sample did not allow for the evolution of more than one sample. Therefore, it was not possible to do a statistical analysis to determine the uncertainties of the observed phenomena and to e.g. follow the evolution of several similar samples.

\section{Results and discussion}
\begin{figure*}[t]
  \centering
  \includegraphics[width=0.7\textwidth]{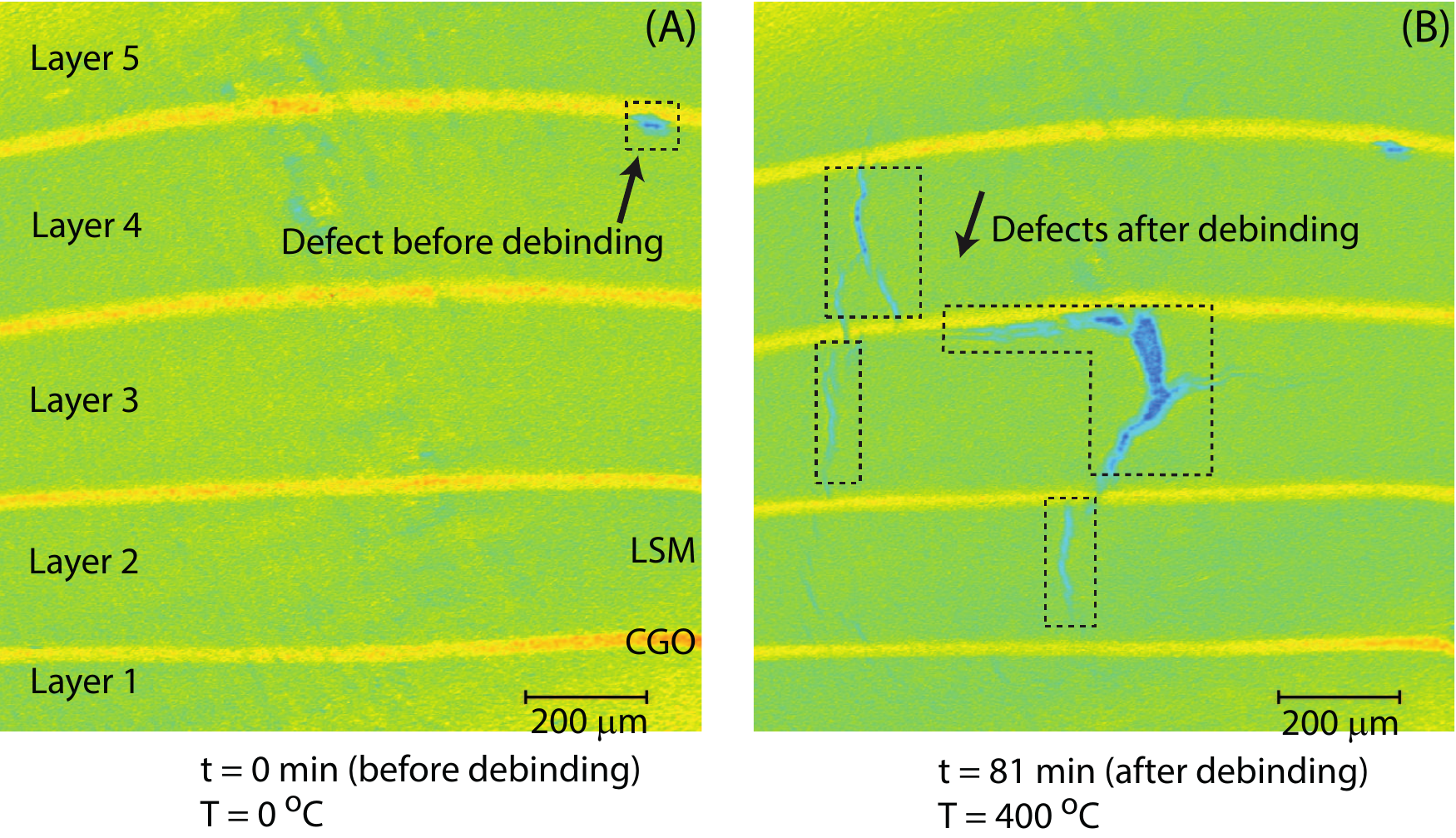}
  \caption{A false-color slice through the sample near the center of the sample as prepared and after the sample has been heated to 400$^\circ$C during the debinding step. The thin CGO layers can clearly be seen. The defect(s) before (A) and after debinding (B) have been indicated, as well as the layer numbers. The non-indicated pores in the LSM layers are due to noise from the reconstruction and a slight variations in porosity in the layer itself.}
  \label{Fig_After_debinding}
\end{figure*}

The evolution of several and diverse flaws can be followed using X-ray tomographic microscopy and their reciprocal influence on the microstructure can be observed. A slice through the reconstructed volume near the center of the sample before and after debinding is shown in Fig. \ref{Fig_After_debinding} for the multilayer sample considered here. Initially the sample contains a single defect as result of the lamination process (probably an air inclusion trapped during the lamination process), as is clearly indicated in the figure. The rest of the material is homogeneous and defect free. During debinding of the sample a crack was formed in the center of the sample, leading to a slight delamination at the middle CGO layer. The cracks formed are mainly vertical and are formed in the LSM layer, which is the one with higher organic content. The debinding process is thus critical for the sample. The critical stress causing the cracks to appear is caused by gas formation from the debinding process. This process step usually requires long time and controlled conditions. However, due to the small geometry of the sample it can be assumed that the defects formed represented an intrinsic weakness of the design chosen. In a production sample these cracks are already critical. It is worth noticing that the initial defect seen in Fig. \ref{Fig_After_debinding} has not enlarged during debinding. The crack can be seen to mostly affect the middle part of the sample, and the sintering behavior of e.g. the lower part of the sample should not initially be affected by the crack.

\subsection{Densification behavior}
\begin{figure}[t]
  \centering
  \includegraphics[width=1\columnwidth]{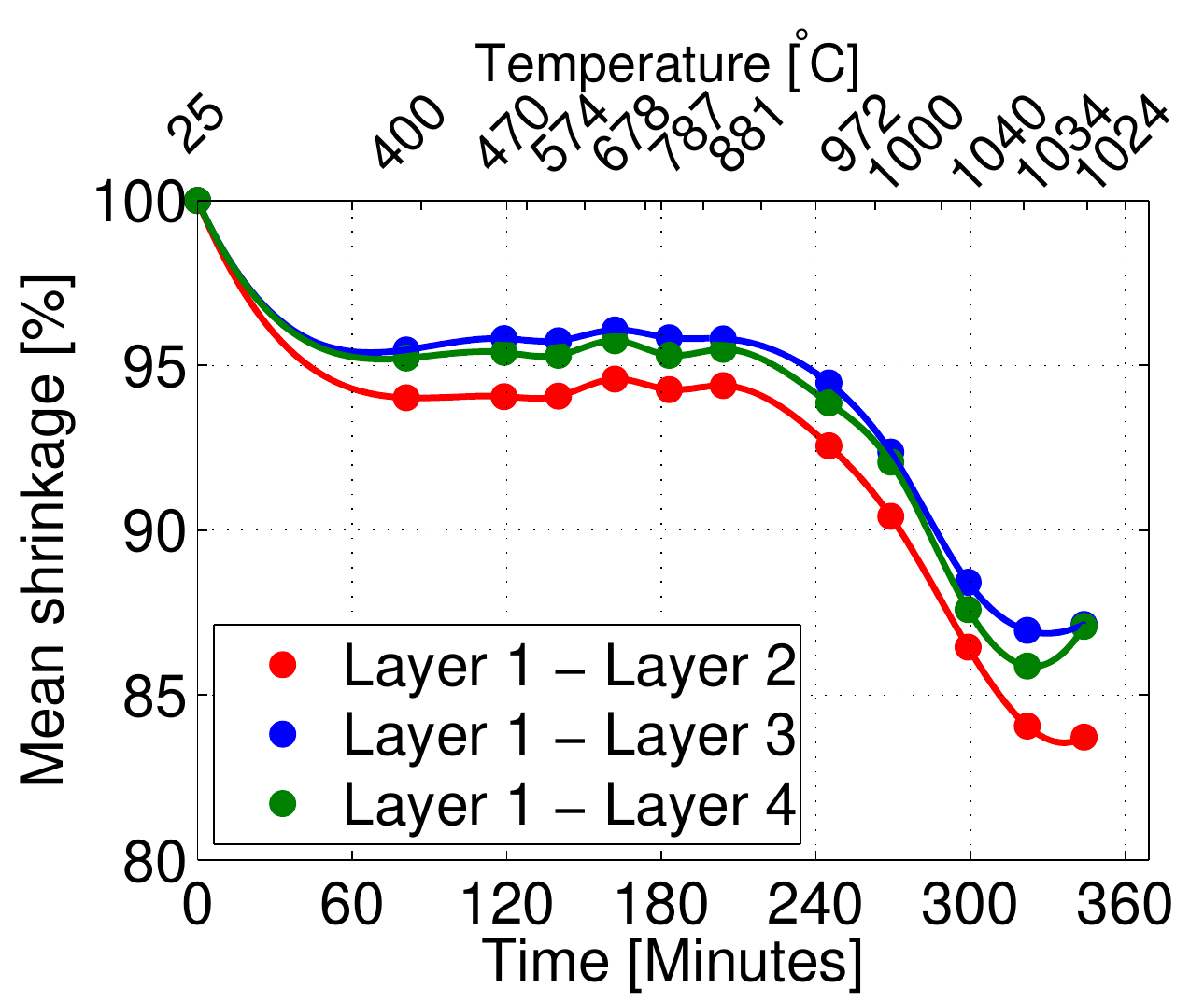}
  \caption{The mean shrinkage in distance between layer one and the three other layers as function of time.}
  \label{Fig_Mean_shrinkage}
\end{figure}

Before considering the evolution of the defects, we first consider the densification behavior of the sample itself. In order to determine this, some features inside the sample, that can be tracked as function of time, are needed. As the sample is oriented vertically, the thin CGO layers can be used as fix points. The shrinkage of the sample in the vertical direction has been found by calculating the average vertical distance between the different CGO layers, as a function of time. This is shown in Fig. \ref{Fig_Mean_shrinkage} for all layers. Here the sample can be seen to contract both during the debinding stage and also as the temperature increases to 1000$^\circ$C and sintering begins. The time resolution was not sufficient to resolve any expansion during the debinding step. However, if the crack formation is caused by gas formation, a slight expansion of the sample in the early debinding stage, followed by a net contraction, could explain the observed phenomena. The shrinkage is seen to be slightly larger for the bottom layers, which is a consequence of these being least affected by the cracks in the sample.  As the sintering temperature is reached, the diffusivity of the materials increases, and thus the sintering stress also increases.

\begin{figure*}
  \centering
\includegraphics[width=1\textwidth]{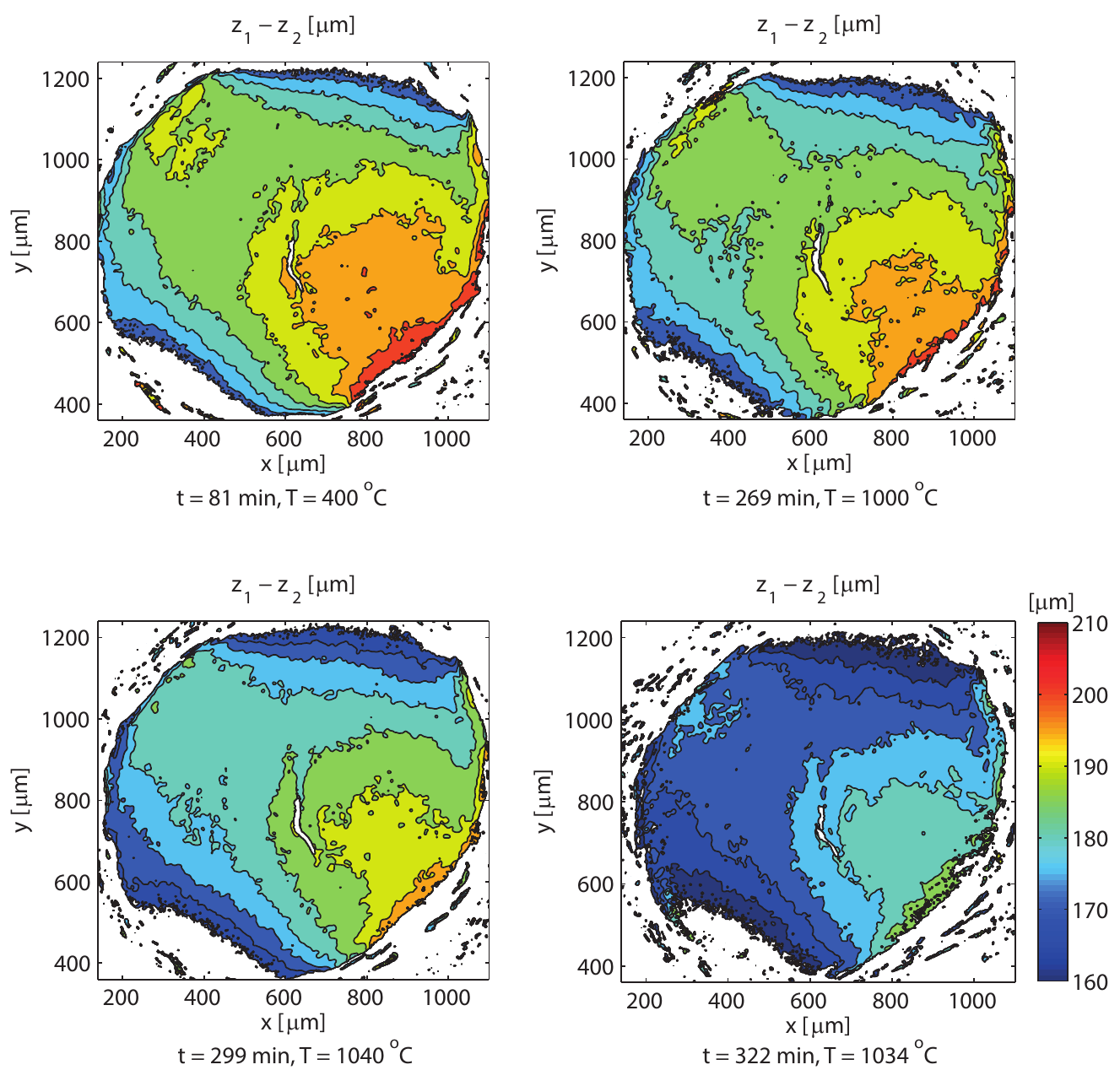}
\caption{The difference in distance between CGO layers one and two at different times, $t$. The sample can clearly be seen to be contracting.}
  \label{Fig_height_layer_1_2}
\end{figure*}

Tomographic microscopy also allows for a more detailed understanding of the sintering behavior. Instead of computing the average distance between the CGO layers, we can compute the vertical distance to the next layer, for each point in the layers. This is shown in Fig. \ref{Fig_height_layer_1_2}, which shows the vertical distance between CGO layer 1 and CGO layer 2 for each point in the layers, at different times. As can clearly be seen the distribution of distance is saddle-formed shaped. Determining this structure with the use of conventional sintering measurements techniques would have been impossible. As time progresses and sintering evolves the distance between the two CGO layers can be seen to shrink. As will be discussed subsequently, after $t$ = 322 min, corresponding to 52 minutes of isothermal treatment, the second layer is so affected by the spreading crack that delamination occurs. Until this time the shrinkage in distance is homogeneous between layer one and two.

\section{Defect evolution}
\begin{figure*}[t]
  \centering
  \includegraphics[width=1\textwidth]{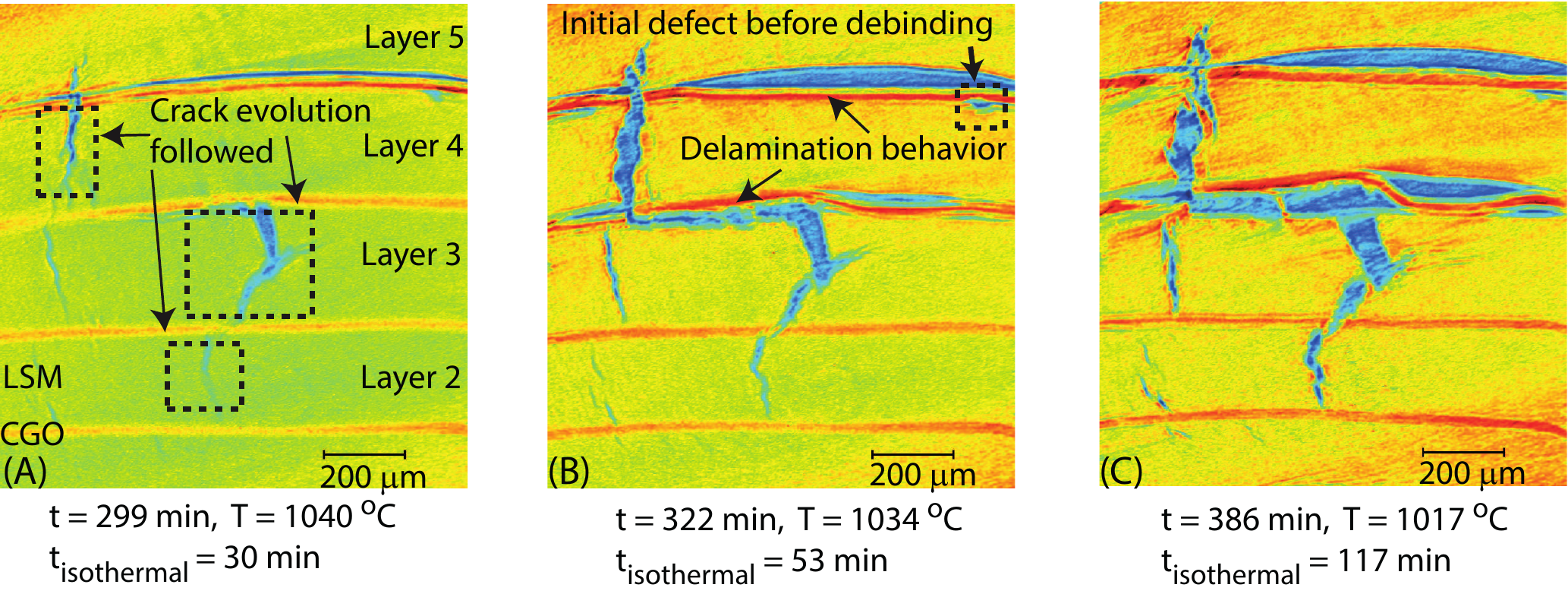}
  \caption{A false-color slice through the sample near the center at the three different time steps of $t$=[299 322 386] min, corresponding to Fig. (A), (B) and (C), respectively. The regions in which the crack growth have been followed, has been indicated by the arrows. As can be seen compared with Fig. \ref{Fig_After_debinding}, substantial delamination has occurred along several of the CGO layers. This happened at the onset of sintering, over a very short time span.}
  \label{Fig_Crack_indications}
\end{figure*}

It is important to consider the evolution of the defects present in the sample, both to study their influence on the sintering behavior but also to determine when their influence becomes critical to the overall stability of the sample. We first consider the temporal evolution of the cracks, apparent in Fig. \ref{Fig_After_debinding} after debinding. These cracks are ``naturally'' occurring cracks, caused by a too rapid debinding cycle and the intrinsic sample geometry. Previously crack growth along interfaces in porous ceramic layers have been studied in great detail, using e.g. in situ observations of crack growth to determine the macroscopic fracture energy \citep{Soerensen_2001}. The evolution of the system analyzed here is too rapid for this kind of analysis, but it is sufficiently slow that the crack growth perpendicular to the layer interface can be studied. Conventionally, fragile facture analysis is performed on a statistical basis. The experiment described here can instead be used to give a general idea on the fracture evolution, and to follow the features of fractures on a level of detail which is not possible using a broad statistical analysis. This detailed analysis has been done for the three thick interlayer LSM layers in the sample that developed cracks, i.e. layers two, three and four, as shown in Fig. \ref{Fig_Crack_indications}. It is clearly seen that the crack size evolution is significantly slower for interlayer cracks than the cracks that propagate along the layer interfaces.

The evolution of the cracks is best illustrated by considering the total volume of the cracks as function of time. The volume of the cracks can be determined based on identifying all voxels with a value lower than a given threshold, here termed $\alpha$. Changing the value of the threshold will of course change the volumes of the cracks. However, we have observed that the slope of the crack volume in time, i.e. the evolution of the size of the crack with time, remain similar for the suitable range of $\alpha$-values used to determine the cracks. Therefore, a fixed threshold has been used to analyse the crack growth data.

\begin{figure}[t]
  \centering
  \includegraphics[width=1\columnwidth]{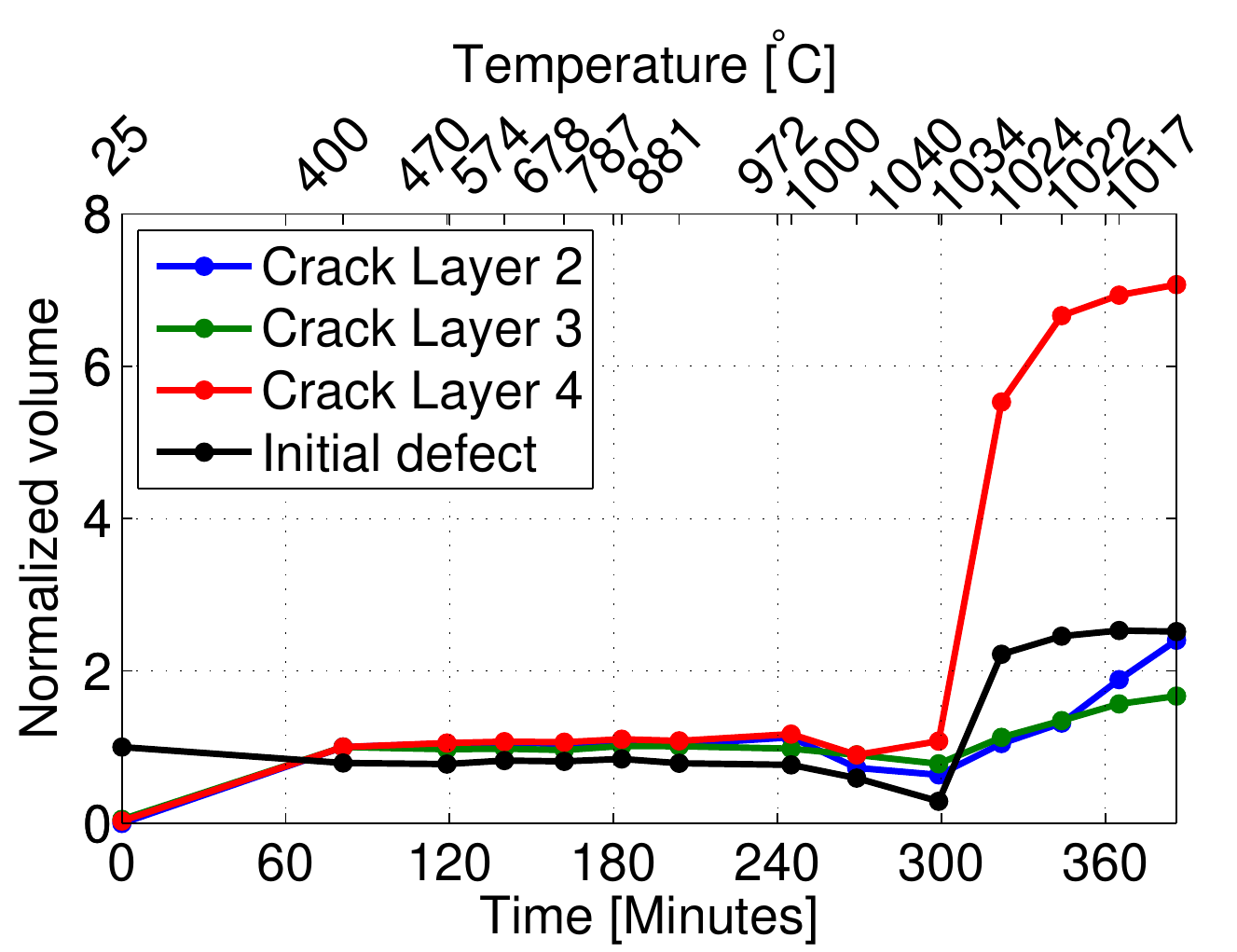}
  \caption{The normalized volume of the interlayer cracks and the initial defects as a function of time. For the cracks the volume has been normalized to the volume after debinding.}
  \label{Fig_Crack_volume_all_layers}
\end{figure}

\begin{figure*}[t]
  \centering
  \includegraphics[width=1\textwidth]{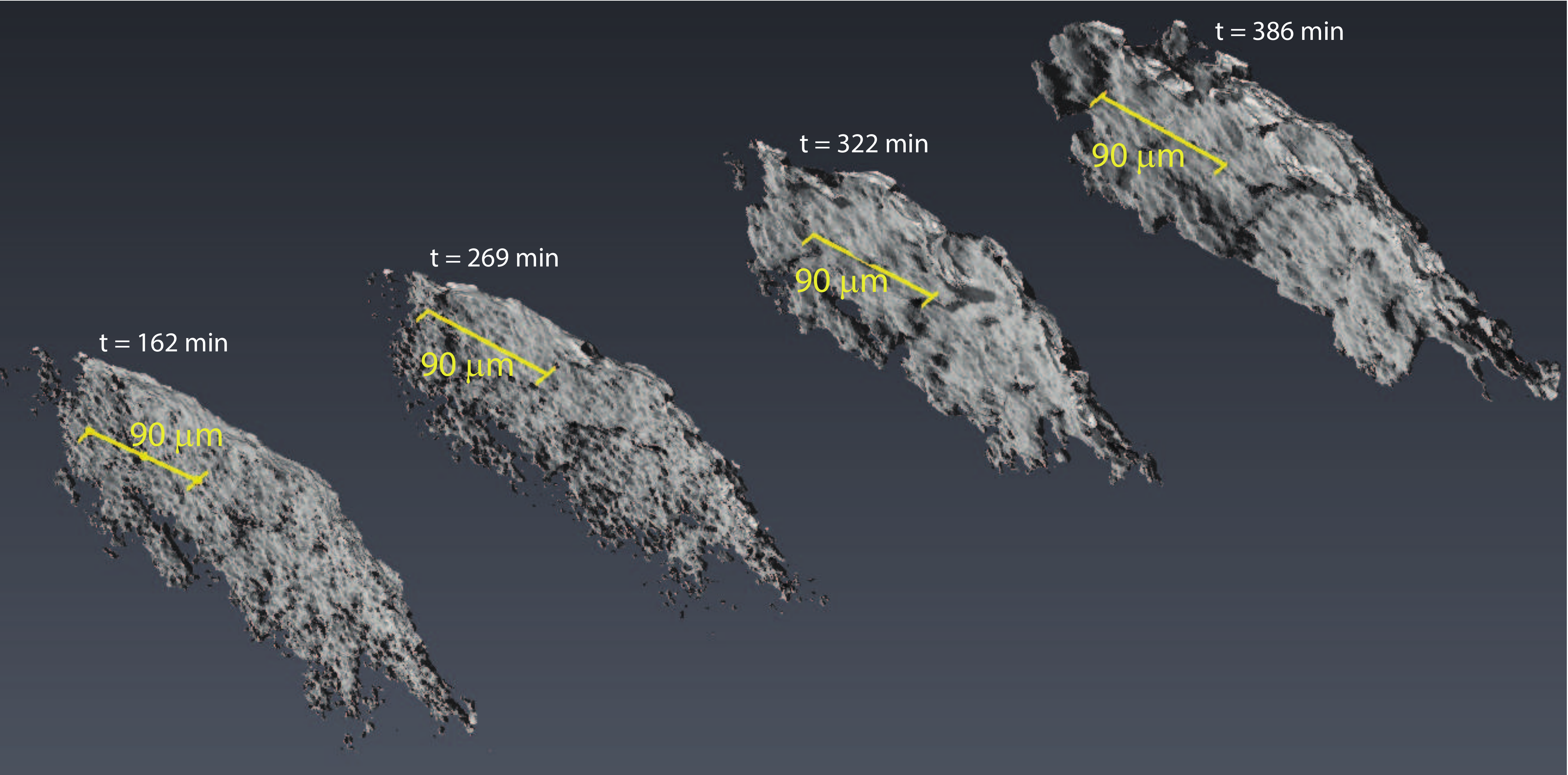}
  \caption{The surface of the crack in layer 3 for selected times. The top of the crack is at the CGO layer between layer 3 and 4.}
  \label{Fig_Isosurfes_layer_2_0.07}
\end{figure*}

Shown in Fig. \ref{Fig_Crack_volume_all_layers} is the normalized volume of the part of the cracks extending into LSM layers 2, 3 and 4. The volume of the crack in the entire sample is not shown, but only the interlayer crack volume, i.e. in the regions indicated in Fig. \ref{Fig_Crack_indications}. The volume of the delaminations are thus excluded. These interlayer regions are selected, as far as possible, to only contain the interlayer cracks. The volume has been normalized to the volume of the crack after debinding. As can be seen from the figure the crack in all layers appear after debinding, and remain constant in size until the sample reaches a temperature where sintering occur (at $t$=245 min). After the onset of sintering the evolution of the cracks are substantially different, at least in terms of size. The cracks in all layers can be seen to shrink slightly at the beginning of sintering, but when sintering begins in earnest (at $t$=299 min) the volume of the crack in layer 4 increases dramatically while the size of the crack in layer 2 and 3 increase at a much slower linear rate. We propose that the difference in behavior is caused by the rupture of the thin CGO layer intermediate between LSM layers 4-5. Since the crack penetrates the thin CGO layer, it is able to open and expand much more than the cracks in layers 3 and 4, where the cracks evolve along the CGO layer instead of rupturing it completely. A closer look at the crack seen in the middle of the sample, i.e. in layer 3, is shown in Fig. \ref{Fig_Isosurfes_layer_2_0.07}, which show isosurfaces of the crack at different times. Here the change in the crack volume can clearly be seen as time increases.

\begin{figure*}[t]
  \centering
  \includegraphics[width=0.6\textwidth]{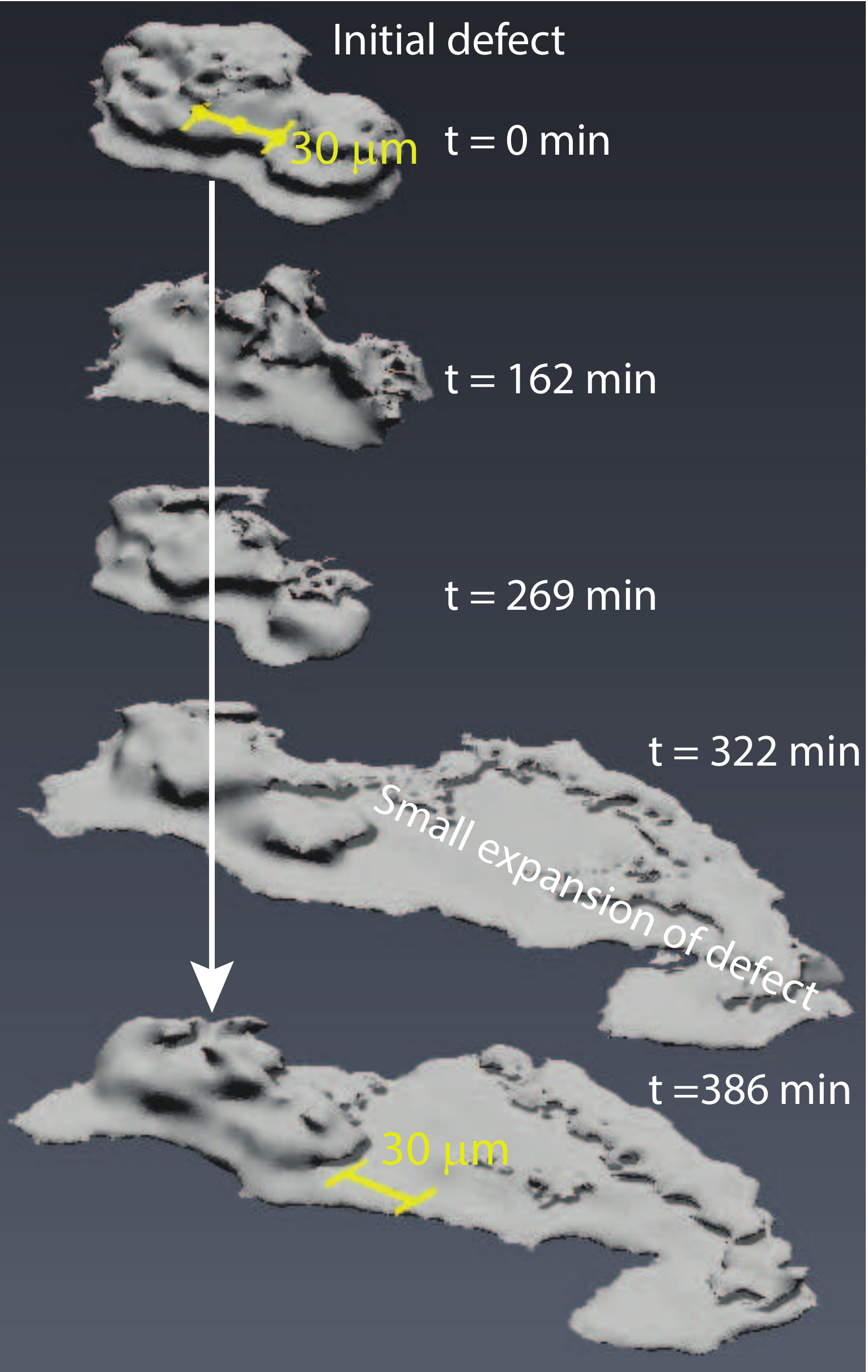}
  \caption{The surface of the initial defect for selected times. The rough shape of the initial defect can be seen to be preserved throughout time. The bottom of the structure is at the CGO layer between layer 4 and 5, and the structure extends into layer 4 from there, i.e. the z-axis is reversed in the image. The image is not to scale compared with Fig. \ref{Fig_Isosurfes_layer_2_0.07}.}
  \label{Fig_Isosurfes_defect_2_0.07}
\end{figure*}

The normalized volume of the initial defect present before debinding, i.e. the defect indicated in Fig. \ref{Fig_After_debinding}(A), is also shown in Fig. \ref{Fig_Crack_volume_all_layers} as a function of time. The evolution can be seen to be somewhat similar to that of the interlayer cracks, albeit with a smaller change in volume compared to the layer 4 crack. It is clear from these considerations that the initial defect does not cause the observed delamination behavior. A series of isosurface images of the initial defect is shown in Fig. \ref{Fig_Isosurfes_defect_2_0.07}. From these it can clearly be seen that while the defect expands along the thin CGO layer, it remains isolated from the other cracks/defects that develop in the sample.

\section{Conclusion}
We have demonstrated that X-ray tomographic microscopy can be successfully used to follow the sintering behavior of a multilayer ceramic sample, and can be used to study sintering, delamination and crack formation and growth during sintering. The results show that naturally occurring defects, caused by the tapecasting process, might not be critical in sample degradation. The debinding step, on the other hand, is where defects are nucleated. It was shown that crack growth along the material layers is significantly faster along the layers rather than perpendicular to them, and that crack growth only accelerates when sintering occurs. Thus sintering is creating the propagation conditions for the crack.

\section*{Acknowledgements}
The authors would like to thank the Danish Council for Independent Research Technology and Production Sciences (FTP) which is part of The Danish Agency for Science, Technology and Innovation (FI) (Project \# 09-072888) for sponsoring the OPTIMAC research work.  The authors also which to thank DanScatt - Danish Centre for the use of Synchrotron X-ray and Neutron facilities for sponsoring the research work presented here. JLF acknowledges CCMX for funding.

\clearpage
\clearpage

\bibliographystyle{elsarticle-harv}

\end{document}